\documentclass{IEEEcsmag}

\usepackage[colorlinks,urlcolor=blue,linkcolor=blue,citecolor=blue]{hyperref}

\usepackage[utf8]{inputenc}
\usepackage{upmath}
\usepackage{mathtools}
\usepackage[style=ieee]{biblatex}
\usepackage{listings}
	\lstdefinelanguage{diff}{
    basicstyle=\ttfamily\small,
    morecomment=[f][\color{diffstart}]{@@},
    morecomment=[f][\color{diffincl}]{+},
    morecomment=[f][\color{diffrem}]{-},
  }
\usepackage[svgnames]{xcolor}
  \definecolor{diffstart}{named}{Grey}
  \definecolor{diffincl}{named}{Green}
  \definecolor{diffrem}{named}{OrangeRed}
\usepackage{xspace}
\let\labelindent\relax
\usepackage[inline]{enumitem}
\usepackage{caption} 
\usepackage{amssymb}
\usepackage{numprint}
\usepackage{tabu}
\usepackage{multirow}

\newcommand{\ie}{\textit{i.e.,}\xspace}
\newcommand{\eg}{\textit{e.g.,}\xspace}
\newcommand{\etc}{\textit{etc.}\xspace}

\newcommand{\listref}[1]{Listing~\ref{#1}\xspace}
\newcommand{\tabref}[1]{Table~\ref{#1}\xspace}

\usepackage{cancel}

\bibliography{references.bib}

\begin{document}

\title{Using Sequence-to-Sequence Learning for Repairing C Vulnerabilities}

\author{Zimin Chen}
\affil{KTH Royal Institue of Technology \\ zimin@kth.se}

\author{Steve Kommrusch}
\affil{Colorado State University \\ steveko@cs.colostate.edu}

\author{Martin Monperrus}
\affil{KTH Royal Institue of Technology \\ martin.monperrus@csc.kth.se}

\begin{abstract}
Software vulnerabilities affect all businesses and research is being done to avoid, detect or repair them. In this article, we contribute a new technique for automatic vulnerability fixing. We present a system that uses the rich software development history that can be found on GitHub to train an AI system that generates patches. We apply sequence-to-sequence learning on a big dataset of code changes and we evaluate the trained system on real world vulnerabilities from the CVE database. The result shows the feasibility of using sequence-to-sequence learning for fixing software vulnerabilities.
\end{abstract}

\maketitle

\chapterinitial{A software vulnerability} is a weakness in code that can be exploited by an attacker to perform unauthorized actions. For example, one common kind of vulnerability is a buffer overflow, which allows an attacker to overwrite a buffer's boundary and inject malicious code. Another example is an SQL injection, where malicious SQL statements are inserted into executable queries. The exploitation of vulnerabilities contributes to the hundreds of billions of dollars that cybercrime costs the world economy each year \cite{losses2014estimating}.

Each month, thousands of such vulnerabilities are reported to the Common Vulnerabilities and Exposures (CVE) database. Each one of them is assigned a unique identifier and category. For instance, the entry identified by CVE-2019-9208 is a vulnerability in Wireshark which causes a null pointer exception, and it is categorized as `Null Pointer Dereference'. In October 2019 alone, 1737 vulnerabilities were reported to the National Vulnerability Database (NVD), which is the main CVE databases. Each vulnerability represents a threat until a patch is written by the developers.

\noindent
\begin{minipage}{\linewidth}
\begin{lstlisting}[language=diff,columns=flexible, frame=single, basicstyle=\ttfamily\small, escapeinside={(*@}{@*)}, label={lst:CVE-example},caption={Patch for CVE-2009-4004 and CVE-2016-8658, the developers added a boundary check in both cases.}, captionpos=b, breaklines=true]
(*@ \hspace{1.5cm} \textbf{Patch for CVE-2009-4004} @*)
    unsigned bank_num = mcg_cap & 0xff, bank;

    r = -EINVAL;
-   if (!bank_num)
+   if (!bank_num || bank_num >= KVM_MAX_MCE_BANKS)
        goto out;
    if (mcg_cap & ~(KVM_MCE_CAP_SUPPORTED | 0xff | 0xff0000))
        goto out;
        
(*@ \hspace{1.5cm} \textbf{Patch for CVE-2016-8658} @*)
        (u8 *)&settings->beacon.head[ie_offset],
        settings->beacon.head_len - ie_offset,
            WLAN_EID_SSID);
-   if (!ssid_ie)
+   if (!ssid_ie || ssid_ie->len > IEEE80211_MAX_SSID_LEN)
        return -EINVAL;

    memcpy(ssid_le.SSID, ssid_ie->data, ssid_ie->len);
\end{lstlisting}
\end{minipage}

There are many available patches fixing a given kind of vulnerability. Let us consider CVE category `Improper Restriction of Operations within the Bounds of a Memory Buffer': CVE-2009-4004 and CVE-2016-8658 are both vulnerabilities in the Linux kernel, published in the NVD in Nov 2009 and Oct 2016 respectively. The corresponding patches to fix them are shown in \listref{lst:CVE-example}. Both vulnerabilities are fixed by adding a check on the memory boundary.

Considering that millions of commits are publicly available on open-source projects, we assume we have enough data to build a system that would learn from past commits, and use it to predict patches for new vulnerabilities. This is the contribution of this article, we present a novel and original approach for learning how to automatically generate patches for security vulnerabilities.

We use a powerful machine learning technique called sequence-to-sequence learning (seq2seq). Seq2seq learns the mapping between a sequence of tokens to another sequence. It is heavily used to automatically translate between different human languages (think Google Translate) and has pushed forward the state-of-the-art performance there. For machine translation, the training data for the seq2seq model consists of a corpus of pairs of sentences, \eg sentences in English and the corresponding translation in French. To train seq2seq for software vulnerability repair, we need such a corpus, but for source code. 

In this work, we have collected a dataset consisting of 21 million bug fix commits on GitHub from 2017 and 2018. The collected dataset is used to train a seq2seq model. To assess the effectiveness of the trained model, we collected real-world CVE vulnerabilities from four well-known open-source projects: Linux kernel, OpenSSL, systemd, and Wireshark. Our results show the promise of using seq2seq for fixing vulnerabilities, with 26.7\%, 13.7\% and 9.2\% accuracy for fixing vulnerable functions of different sizes. 

Notably, our approach is fully generic, instead of focusing on one particular kind of vulnerability such as buffer overflow \cite{gao2016bovinspector}. Compared to the related work, we evaluate our technique on a large number of real vulnerabilities found in CVE databases, not on synthetic code \cite{harer2018learning} or on manually collected benchmark \cite{ma2017vurle}. To our knowledge, our experiment involves the largest ever training dataset for AI on code. 

In summary, our contributions are:
\begin{itemize}
\item We mine two years of bug fixes from GitHub, and we share the largest dataset for machine learning on code changes (21 million bug fix commits with \numprint{910000} commits related to C).
\item We successfully train a seq2seq model using the byte pair encoding to address the size of the vocabulary in source code, which reaches 6 million different words in our dataset.
\item We report original results on real-world vulnerability fixing with machine learning, 14/630 vulnerabilities with CVE identifiers can be fixed in a fully automated manner.
\end{itemize}

\section{Sequence-to-sequence learning}
Sequence-to-sequence (seq2seq) learning is a modern machine learning framework that is used to learn the mapping between two sequences, typically of words \cite{sutskever2014sequence}. It is widely used in automated translation, text summarization and other tasks related to natural language. A seq2seq model consists of two parts, an encoder and a decoder. The encoder maps the input sequence $X = (x_{0},x_{1},...,x_{n})$ to an intermediate continuous representation $H = (h_{0},h_{1},...,h_{n})$. Then, given $H$, the decoder generates the output sequence $Y = (y_{0}, y_{1},...,y_{m})$. Note that the size of the input and output sequences, $n$ and $m$, can be different. A seq2seq model is optimized on a training dataset to maximize the conditional probability of $p(Y \mid X)$, which is equivalent to:

\begin{align*}
\begin{split}
p(Y \mid X) &= p(y_{0},y_{1},...,y_{m} \mid x_{0},x_{1},...,x_{n})\\
            &= \prod_{i=0}^{m} p(y_{i} \mid H, y_{0},y_{1},...,y_{i-1})
\end{split}
\end{align*}

Prior work has shown that source code is as natural as human language \cite{hindle2012naturalness}, and techniques used in natural language processing have been demonstrated to work well on source code, including seq2seq learning \cite{chen2019sequencer}. In our work, we use a seq2seq model called ``transformer'' \cite{vaswani2017attention}. The transformer model is the state-of-the-art architecture for seq2seq learning. 

\section{Rare words in source code}
One of the main challenges of using the seq2seq model on source code is that it hardly handles very rare words \cite{hellendoorn2017deep}. The problem is that rare words, such as original literals or domain-specific identifiers, are too uncommon or even non-existent in the collected training data, and hence cannot be used when decoding. Indeed, in source code, rare variable and function names are more common compared to human language. A basic technique to handle the rare word problem is to increase the vocabulary size, say from 10k to 50k, but this is only a workaround: there will always be rare words for which not enough data is available at training time.

However, a rare word may have subwords that are frequent. For example the word \textit{underworld} might be a rare word, but the subwords, \textit{under} and \textit{world} are common words. So if we could represent our vocabulary with frequent subwords, then we could generate any word with it. Byte pair encoding (BPE) is the state-of-the-art technique for learning the most frequent subwords \cite{sennrich2015neural}. BPE starts with the basic characters as the vocabulary (\eg the letters in the Latin alphabet). At each time step, the most frequent subword pair is combined into one new subword, and it is added into the vocabulary. It continues until we have reached a predefined vocabulary size. \listref{lst:bpe-example} shows an example of applying BPE on a C function. Variables like \textit{destroyKeyValuePair} and \textit{freeValue} are considered as respectively 4 subwords and 2 subwords (\textit{destroy Key Value Pair} and \textit{free Value}), which are more common words that can be expected to be found in other code snippets.
BPE has been successfully used in machine translation \cite{sennrich2015neural} and source code modeling \cite{karampatsis2019maybe}. In this paper, we are the first to report on using seq2seq with BPE for patch generation.

\noindent
\begin{minipage}{\linewidth}
\begin{lstlisting}[columns=flexible, frame=single, captionpos=b, basicstyle=\ttfamily\small, escapeinside={(*@}{@*)}, breaklines=true, label={lst:bpe-example}, caption={Example of applying learned BPE on a C function. ''\_''(U+2581) indicates the start of a new word. BPE learned some useful subwords such as Key, Value and Pair.}]
(*@ \hspace{2.5cm} \textbf{C code} @*)
void destroyKeyValuePair(keyValuePair kvPair) { 
    kvPair -> freeValue(kvPair -> value); 
    kvPair -> freeKey(kvPair -> key); 
    free(kvPair); 
}
(*@ \hspace{1.5cm} \textbf{After applying BPE} @*)
_void _destroy Key Value Pair _( _key 
Value Pair _kv Pair _) _{ _kv Pair _->
_free Value _( _kv Pair _-> _value _) _; 
_kv Pair _-> _free Key _( _kv Pair _-> 
_key _) _; _free _( _kv Pair _) _; _}
\end{lstlisting}
\end{minipage}

\section{Data collection and filtering}

\textbf{Training dataset.} To train a seq2seq model, we need a large corpus of buggy and fixed source code. We create such a corpus by mining the GitHub development platform. We use GH Archive \cite{GHArchive} to download all GitHub events that happened between 2017-01-01 and 2018-12-31. These events can be triggered by a Github issue creation, an opened pull request, and other development activities. In our case, we focus on push events, which are triggered when a commit is pushed to a repository branch. To only collect bug fix commits, we adopt a keyword-based heuristic \cite{Martinez2013}: if the commit message contains keywords (\textit{fix} OR \textit{solve} OR \textit{repair}) AND (\textit{bug} OR \textit{issue} OR \textit{problem} OR \textit{error} OR \textit{fault} OR \textit{vulnerability}), we consider it a bug fix commit and add it to our corpus. In total, we have analyzed 730 million commits and selected 21 million commits identified as bug fix commits. 

In our experiment, we focus on C code as the target programming language for automatic repair. Therefore we further filter the bug fix commits based on the file extension. We remove commits that did not fix any file that ends with '.c', this leaves us with \numprint{910000} buggy C commits. Then, for each commit, we extract function pairs that were changed in the commit. We are learning function-level changes instead of file-level changes because seq2seq suffers from long input and output \cite{cho2014properties}. To identify function-level changes, we use the GNU compiler preprocessor to remove all comments, and we only extract functions that are changed. Then, we used Clang to parse and tokenize the function source code.

In the end, we obtain \numprint{1806879} function-level changes, reduced to \numprint{642399} after removing duplicates. The sizes of functions vary and we observe some C functions are still too big to be learned with seq2seq. Therefore, we further divide the training dataset to $d_{200}$, $d_{100}$, and $d_{50}$, where the function lengths in before and after the change are limited to 200, 100 and 50 tokens respectively. In $d_{200}$, $d_{100}$, and $d_{50}$, we have \numprint{299976}, \numprint{146051}, and \numprint{49340} function-level changes respectively. The function code before the change is used as input to the seq2seq model, and the function after the change is used as the ground truth output for training.

\textbf{Testing dataset.} We also collect a dataset for testing the ability of seq2seq to fix real vulnerabilities. 
We used Data7 \cite{JimenezPT18} to collect known vulnerabilities with CVE identifier from four well-known projects: Linux kernel, OpenSSL, systemd, and Wireshark. Each sample in the testing dataset consists of a CVE number and a list of commits that fixed the vulnerability. Next, we extract function-level changes from these vulnerabilities, and we call them vulnerable functions. We consider a vulnerability to be completely fixed if all its vulnerable functions are fixed. A vulnerability is partly fixed if at least one of its vulnerable functions is fixed. Test sets $t_{200}$, $t_{100}$, and $t_{50}$ are created by including only vulnerable functions where the token lengths before and after the change are limited to 200, 100, and 50 tokens respectively. For $t_{200}$, we have 1615 vulnerable functions, that represent 630 vulnerabilities. For $t_{100}$, we have 725 vulnerable functions, that represent 288 vulnerabilities. For $t_{50}$, we have 120 vulnerable functions spread over 85 vulnerabilities.

\section{Experiment setup}
The training datasets $d_{200}$, $d_{100}$, and $d_{50}$ are  randomly divided into training data and validation data, with 98\% as training and 2\% as validation. We select the best models with the highest validation accuracy a grid search in the hyper-parameter space. We evaluate the resulting models on our testing datasets, $t_{200}$, $t_{100}$ and $t_{50}$.
We train three baseline seq2seq models on the three datasets $d_{200}$, $d_{100}$ and $d_{50}$ with a vocabulary set to the top 50k most common tokens. Those baselines represent the state-of-the-art seq2seq model, without specific care to address the rare word problem.

Next, we explore seq2seq models using BPE for handling rare tokens. For the BPE configuration, we set the size of subword vocabularies to either 1000, 5000 or 10000, \ie the vocabulary is the top 1000, 5000 or 10000 most frequent subwords in source code identifiers. After having identifier the optimal BPE subvocabulary, we traing the seq2seq models our training dataset $d_{200}$, $d_{100}$ and $d_{50}$. Consequently, in addition to our baselines, we have nine different settings: the cross-product of three different token length limits and the three different vocabularies defined by BPE. 
In total, we have 12 different seq2seq models summarized in \tabref{tab:performance}. The seq2seq model are called after the training dataset and the BPE configuration:
for example $BPE_{1000} - d_{200}$ refers to the seq2seq model trained on $d_{200}$ and with vocabulary set to the top 1k most common subwords.

The best model for all 12 different settings is evaluated on the corresponding testing dataset: $Baseline - d_{200}$ is evaluated on $t_{200}$, $BPE_{1000} - d_{100}$ is evaluated on $t_{100}$, \etc
We use beam search to predict fixes of vulnerable functions which means that the seq2seq model generates the top 50 most likely predictions per vulnerable function. The vulnerable function is considered as fixed when one of 50 predictions matches the ground truth human fix, as done in prior work \cite{chen2019sequencer, tufano2018empirical}.

We use the OpenNMT-tf \cite{2017opennmt} for training the transformer model, and SentencePiece \cite{kudo2018sentencepiece} for learning BPE on the training data.

\section{Experimental Results}

\begin{table*}
\begin{center}
\begin{tabu}{|c|[2pt]c|[2pt]c|[2pt]c|}
\hline
\multirow{2}{*}{Model} & \multirow{2}{*}{Fixed functions} &  \multicolumn{2}{c|}{Vulnerabilities} \\ \cline{3-4}
& & \multicolumn{1}{c|}{Partially fixed} & \multicolumn{1}{c|}{Completely fixed} \\ \tabucline[2pt]{-}
$Baseline - d_{50}$ & 5/120 (4.2\%) & 3/85 (3.5\%) & 1/85 (1.2\%) \\ \cline{2-4}
$BPE_{1000} - d_{50}$ & 26/120 (21.7\%) & 17/85 (20\%) & \textbf{3/85 (3.5\%)} \\ \cline{2-4}
$BPE_{5000} - d_{50}$ & \textbf{32/120 (26.7\%)} & \textbf{22/85 (25.9\%)} & \textbf{3/85 (3.5\%)} \\ \cline{2-4}
$BPE_{10000} - d_{50}$ & 28/120 (23.3\%) & 18/85 (21.1\%) & \textbf{3/85 (3.5\%)} \\ \tabucline[2pt]{-}
$Baseline - d_{100}$ & 2/725 (0.3\%) & 2/288 (0.7\%) & 0/288 (0\%) \\ \cline{2-4}
$BPE_{1000} - d_{100}$ & 68/725 (9.4\%) & 40/288 (13.9\%) & 6/288 (2.1\%) \\ \cline{2-4}
$BPE_{5000} - d_{100}$ & 97/725 (13.4\%) & \textbf{47/288 (16.3\%)} & 5/288 (1.7\%)\\ \cline{2-4}
$BPE_{10000} - d_{100}$ & \textbf{99/725 (13.7\%)} & 45/288 (15.6\%) & \textbf{10/288 (3.5\%)} \\ \tabucline[2pt]{-}
$Baseline - d_{200}$ & 0/1615 (0\%) & 0/630 (0\%) & 0/630 (0\%) \\ \cline{2-4}
$BPE_{1000} - d_{200}$ & 109/1615 (6.7\%) & 45/630 (7.1\%) & 9/630 (1.4\%) \\ \cline{2-4}
$BPE_{5000} - d_{200}$ & 131/1615 (8.1\%) & 52/630 (8.3\%) &  9/630 (1.4\%) \\ \cline{2-4}
$BPE_{10000} - d_{200}$ & \textbf{148/1615 (9.2\%)} & \textbf{55/630 (8.7\%)} &  \textbf{14/630 (2.2\%)} \\ \hline
\end{tabu}
\caption{Performance of our trained seq2seq models on the testing datasets. The first column gives the kind of seq2seq model. The second column shows the accuracy on $t_{50}$, $t_{100}$ and $t_{200}$ respectively. The third column displays the number of partly fixed vulnerabilities. The fourth column represents the number of completely fixed vulnerabilities.}
\label{tab:performance}
\end{center}
\end{table*}

Can the trained seq2seq models generate patches for real-world vulnerabilities?
The main results are presented in \tabref{tab:performance}. The first column gives the name of seq2seq model depending on its BPE configuration. The second column shows the prediction accuracy on $t_{50}$, $t_{100}$ and $t_{200}$ respectively. The third column displays the number of partly fixed vulnerabilities. The fourth column represents the number of completely fixed vulnerabilities.

We first focus on the number of C functions with vulnerabilities that are correctly patched by our system.
Overall, our model is able to fix real world vulnerabilities, up to 32/120 (26\%) for vulnerabilities in small C functions of less than 50 tokens.
The seq2seq model's performance decreases with the input and output length: the values on the $d_{50}$ models (first group of four rows) are higher than those for the bigger functions: it is easier to fix vulnerabilities in shorter C functions. 

Recall that the \textit{Baseline} models do not use BPE, and \tabref{tab:performance} indicates that their accuracies are close to 0. This shows that a standard seq2seq model with a fixed vocabulary is not an option for handling the rare token problem. To further analyze this phenomenon, we analyzed the \numprint{80750} predictions generated by $Baseline - d_{200}$ (\numprint{1615} vulnerable functions in $t_{200}$ $\times$ 50 predictions per function). We found that \numprint{80047} / \numprint{80346} (99\%) predictions contain out-of-vocabulary tokens, this further confirms the prevalence of rare words in source code. 
Now, our results show that byte-pair encoding (BPE) is a powerful solution to this problem: the number of fixed C functions jumps from 5 to 32 for small functions ($d_{50}$) and from 0 to 148 for larger functions ($d_{200}$). Our data suggests that for large functions, there is still some room for improvement with large subword vocabulary (a subword vocabulary of 20000 would likely increase the accuracy).

Recall that a single vulnerability is often fixed in multiple functions at once. The third and fourth columns of \tabref{tab:performance} focus on the number of fixed vulnerabilities instead of the number of fixed vulnerable functions.
Our seq2seq models are able to partially fix up to 22/85 vulnerabilities in small functions.
Completely fixing a vulnerability is much harder, because all the functions inside the vulnerability must be fixed: for instance in CVE-2011-1771, the fix is done over 95 vulnerable functions. Despite this strong requirement, our approach is able to completely fix 3/85 vulnerabilities in $t_{50}$, 10/288 vulnerabilities in $t_{100}$, and 14/630 vulnerabilities in $t_{200}$. To our knowledge, this is the first report of result with seq2seq learning on fixing general vulnerabilities.

The results of seq2seq models are computed based on the ground truth human fix. In production, such a vulnerability fixing system would be used without a ground truth fix. Based on the suspicious functions, we would filter the output of seq2seq using additional checks: such as compilation (remove uncompilable code) and test execution (remove patches yielding test failures). Previous work has shown that using these automatic filtering techniques correctly filters out up to 97\% of the patches generated with seq2seq \cite{chen2019sequencer}.

\section{Case studies}

\noindent
\begin{minipage}{\linewidth}
\begin{lstlisting}[language=diff,columns=flexible, frame=single, basicstyle=\ttfamily\small,label={lst:CVE-2011-1771},caption={CVE-2011-1771 is successfully fixed by seq2seq.},captionpos=b,breaklines=true]
int cifs_close(struct inode *inode, struct file *file)
{
-   cifsFileInfo_put(file->private_data);
-   file->private_data = NULL;
+   if (file->private_data != NULL) {
+       cifsFileInfo_put(file->private_data);
+       file->private_data = NULL;
+   }

    /* return code from the ->release op is always ignored */
    return 0;
\end{lstlisting}
\end{minipage}

Let us now discuss interesting cases.
CVE-2011-1771 is a vulnerability from the Linux kernel, the description of this vulnerability from NVD is: \textit{The cifs\_close function in fs/cifs/file.c in the Linux kernel before 2.6.39 allows local users to cause a denial of service (NULL pointer dereference and BUG) or possibly have unspecified other impact by setting the O\_DIRECT flag during an attempt to open a file on a CIFS filesystem}. The human patch for this vulnerability is shown in \listref{lst:CVE-2011-1771}. The fix consists of adding a null check for variable  private\_data. Our seq2seq approach is able to generate this exact patch (4 out of 12 models can do so: $BPE_{1000} - d_{50}$, $BPE_{10000} - d_{100}$, $BPE_{1000} - d_{100}$ and $BPE_{1000} - d_{200}$).

\noindent
\begin{minipage}{\linewidth}
\begin{lstlisting}[language=diff,columns=flexible, frame=single, basicstyle=\ttfamily\small,label={lst:CVE-2017-8925},caption={CVE-2017-8925 is successfully fixed by seq2seq.},captionpos=b,breaklines=true]
static int omninet_open(struct tty_struct *tty, struct usb_serial_port *port)
{
-   struct usb_serial	*serial = port->serial;
-   struct usb_serial_port	*wport;
-   
-   wport = serial->port[1];
-   tty_port_tty_set(&wport->port, tty);
-
    return usb_serial_generic_open(tty, port);
}
\end{lstlisting}
\end{minipage}

CVE-2017-8925 is another vulnerability from the Linux kernel, the description from NVD is: \textit{The omninet\_open function in drivers/usb/serial/omninet.c in the Linux kernel before 4.10.4 allows local users to cause a denial of service (tty exhaustion) by leveraging reference count mishandling}. It is categorized as `Improper Resource Shutdown or Release'. The human fix shown in \listref{lst:CVE-2017-8925} removes statements that improperly handle variables `port' and `tty'. This exact patch could be generated by all our seq2seq models trained on $d_{100}$ and $d_{200}$. With appropriate training data, our seq2seq approach to generate vulnerability fixes is able to predict the same patches as human developers.

\section{Conclusion}
Software vulnerabilities are common and can cause much damage. In this paper, we took a step towards automatic repair of security vulnerabilities.
We devised, implemented and evaluated a novel system based on sequence-to-sequence learning on past commits from software repositories. We mined 2 years of commit history from GitHub, and we solved the rare word problem in source code by using the byte-pair encoding technique. Our original results show that real world vulnerable C functions can be fixed in a fully automated, data-driven manner. Future work is required to increase the performance of automatic vulnerability tools on fixing general vulnerabilities, and to explore the integration of such technology into the software development process.

\printbibliography

\end{document}